# Automated Analysis of Topic-Actor Networks on Twitter:

## New approach to the analysis of socio-semantic networks

Iina Hellsten[a] and Loet Leydesdorff[b]

**Abstract**

Social-media data provides increasing opportunities for automated analysis of large sets of textual documents. So far, automated tools have been developed to account for either the social networks between the participants of the debates, or to analyze the content of those debates. Less attention has been paid to mapping co-occurring actors (participants) and topics (content) in online debates that form socio-semantic networks. We propose a new, automated approach that uses a whole matrix approach of co-addressed topics and the actors. We show the advantages of the new approach with the analysis of a large set of English-language Twitter messages at the Rio+20 meeting, in June 2012 (72,077 tweets), and a smaller data set of Dutch-language Twitter messages on bird flu related to poultry farming in 2015-2017 (2,139 tweets). We discuss the theoretical, methodological and substantive implications of our approach, also for the analysis of other social-media data.

[a] University of Amsterdam, Amsterdam School of Communication Research (ASCoR), Nieuwe Achtergracht 166, 1018 WV Amsterdam; e-mail: I.R.Hellsten@uva.nl; mobile +31 6 15463912
[b] University of Amsterdam, Amsterdam School of Communication Research (ASCoR), PO Box 15793, 1001 NG Amsterdam; e-mail: loet@leydesdorff.net

**Introduction**

Social-media data provides social scientists with large textual corpora of complex social interactions in online debates. So far, quantitative methods and automated tools have been developed in two separate strands of network research. On the one side, in social-network analysis, the focus has been on networks of actors, and mapping the relations and structure of social interactions (Borgatti & Foster, 1993; Wasserman & Faust, 1994). On the other side, semantic network mapping has been used for analyzing the content of these messages. Content has been mapped in terms of patterns of co-occurring words (Danowski, 2012; Diesner, 2013), topics detected on the basis of clusters in word co-occurrence networks (e.g., Courtial, 1994; Carley & Kaufer, 1993; Danowski, 2012; Diesner, 2013; Leydesdorff, 1989 and 1991), and implicit frames reflecting latent structures in word (co-)occurrences (Leydesdorff & Hellsten, 2005; Hellsten, Dawson & Leydesdorff, 2010).

Both approaches —social-network analysis and semantic-network analysis—provide partial views on the social-media communications. The challenge of analyzing the co-occurrences of actors and topics in debates requires combining ideas from social and semantic-network analysis, and improving the methods to account for the resulting socio-semantic networks. We propose an alternative approach to mapping actor-topic networks using a "whole matrix" approach, and discuss the relative merits of this approach as compared to the established 2-mode network analysis approach in social-network analysis (Borgatti & Foster, 1997). Our approach is innovative both in terms of the network methods, and its theoretical focus on mapping socio-semantic networks.

First, in terms of methods, we improve on the 2-mode matrix approach that is prominent in social-network analysis (Everett & Borgatti, 2013). We propose to take into account the whole matrix of topics and actors, and show the advantages of our approach as providing more informative results. Second, our contribution to the information and communication sciences is that we shift the focus from social actors and their semantics into co-addressing both actors and topics as inspired by actor-network theory (Latour, 1996). This shift in focus, opens up new avenues for theory-building in the social sciences that is less focused on social actors, and more on addressing actors and topics in communications.

Substantially, our focus is on Twitter messages, and we map the co-occurrences of hashtags (as representation of topics) and usernames (as addressed actors). Furthermore, we show an extension to a 3-mode approach that uses three different types of nodes (authors, actors and topics) in one visualization. In summary, instead of asking who (which author) used which concepts (topics), we ask how actors and topics are co-addressed in communication. This research question builds upon earlier calls for combining actors and topics in actor-network-theory, on the one hand, and semantic and socio-semantic-network analysis, on the other.

**Theoretical framework: Network approach**

The aim of actor-network theory (ANT) that was developed in the social-studies-of-science tradition from the early 1980s onwards, is on developing a relational perspective of social interactions between both humans and non-humans. In the semiotic tradition, both semantics and social relations are considered as 'actants' (Callon & Latour, 1981; Latour, 1987; 1996). Actants can be human or non-human agents (Callon *et al.,* 1986). In addition to the idea of human and non-human actants, ANT theorizes networks as an encompassing relational and dynamic social theory. As opposed to social-network analysis that focuses on interactions of human actors, ANT aims to "follow how a given element becomes strategic through the number of connections it commands, and how it loses its importance when losing its connections" (Latour, 1995, p. 372).

Our approach, indeed, focuses on the power of connections in the social-media debates instead of the social actors in the debates. In brief, we ask who was co-addressed with which topics, instead of who addressed which topics. In the following we call social actors originating communications as 'authors' and actors addressed in the communications as 'actors' whereas we refer to co-addressed topics and actors as 'actants' following the actor-network terminology.

In order to position our approach in relation to the wider network theory, we discuss two strands of network analysis. These two strands—social-network analysis and semantic, co-word analysis—have been developed mainly at arm's length from each other. The challenge of theorizing on meaningful socio-semantic networks and how they could change or enrich empirical research in the information sciences and communication studies, has remained an open question.

In social-network analysis, the methodology to measure interactions among social actors as 'authors' has been elaborated during a number of decades (Wasserman & Faust, 1994). Computer programs make it possible to identify important authors in terms of their centrality in the social networks. In social-network analysis, one has studied 2-mode matrices of social authors and their relations to each other are used in addition to 1-mode matrices of social authors and their attributes (Borgatti & Everett, 1997). However, this methodology does not give access to what specific authors were discussing about, i.e., the content of their communications.

The content of communication has been the subject of semantic-network analysis (Landauer, Foltz, & Laham, 1998) that has attracted growing scholarly attention since the early 1990s (Leydesdorff, 1989, 1991), in particular, in two distinct traditions – one thriving on human or computer-assisted coding, the second applying automated analysis to semantic co-word maps.

Carley and Kaufer (1993), for example, called attention to combining the research fields focusing on symbols and the semantic-network analysis, arguing that these two fields were in need of cross-fertilization. In this approach concepts were defined as the nodes in the network, and density was considered as the frequency of co-occurrences of the concepts in the network. Later on, this approach has evolved into established and systematic research on structures of concept networks using dedicated software packages (e.g., AutoMap and ORA), that are based on coding of words in the text into categories that can include, for example, individual names, organization names, and other relevant categories (Diesner, 2013). This approach to semantic-network analysis, requires data cleaning and manual, or vocabulary-assisted, coding of the texts. After the coding, the software can be used for automated network analysis of (large) sets of texts (Pfeffer & Carley, 2012). This approach extends upon content analysis.

In traditional content analysis (e.g. Krippendorff, 1989), the focus is on explicit frames that are created *ex ante* by the coders when designing a coding scheme. Subsequently, the resulting networks of concepts represent the coders' interpretation of significant concepts instead of implicit or emerging meanings in the texts. In principle, such social-science inspired text analysis is very similar to quantitative methods developed in language studies, such as cognitive linguistics, in which one combines text structure and discourse (Sanders & Spooren, 2010).

Recently, automated analysis has been applied to both content analysis and in semantic-network analysis. Automated content analysis focuses on extracting associative frames of manually constructed actors and issues in text documents (e.g. Schultz *et al.,* 2012), and using automated cluster and sentiment analysis (e.g. Burscher, Vliegenthart & de Vreese, 2015). Factor analysis has been used for automated analysis of topics using word/document matrix (Leydesdorff & Welbers, 2011; Vlieger & Leydesdorff, 2011). This factor-analysis based approach is comparable to topic modelling that uses word distributions to detect topics, and assigning words belonging to specific topics. Most notably using Latent Dirichlet Allocation (LDA) that assigns words into clusters using probability distributions (Blei, Ng & Jordan, 2003). The method has been applied to the analysis of large sets of documents (e.g. Jacobi, van Atteveldt & Welbers, 2016).

In another strand of network semantics, Leydesdorff and Hellsten (2005; 2006) developed automated semantic co-word maps to uncover the implicit frames in text documents without human coding. This so-called vector-space model of mapping words is based on the word-document matrices (Salton & McGill, 1983; Turney & Pantel, 2010). In the semantic maps approach, context is defined as a specific topic under discussion, such as 'genetically modified foods'. Using the word/document matrix, one takes into account not only dyads of co-occurring words, but also single words, triads, etc. Furthermore, the method is able to take into account the relations of co-occurring words and the positions of such words in a set of documents (Leydesdorff & Hellsten, 2005). Our approach builds upon the tools designed for semantic-network analysis of co-occurring words that is suitable for identifying implicit semantic relations in large sets of text documents. The rise of the semantic-network approaches has benefitted from the availability of large scale textual data on social interactions at the Web.

Recently, the results of topic-modelling and the co-word approach have been compared using the same sample data (Leydesdorff & Nerghes, 2017). The results show that topic modelling and the semantic-network approach differ significantly from each other; co-word maps outperform topic models in terms of meaningful clustering of the words using medium-size data sets of up to thousand documents (Leydesdorff & Nerghes, 2017). Larger sets are difficult to validate by human reading (Hecking & Leydesdorff, in preparation).

The contribution of this study is to provide a methodological approach that provides an automated analysis of co-addressing actors and topics in text documents, and can be widely applied to socio-semantic-network analysis. Socio-semantic-network analysis has previously been used to map knowledge networks in science by combining citation networks of academic authors and their semantics by Roth and Cointet (2010), Roth (2013) and Taramasco, Cointet and Roth (2011), and to connect individual authors to the concepts they use, in small-group settings. (Saint-Charles & Mongeau, 2018; Basov, Lee & Antoniuk, 2017). We argue that focusing on co-addressing of actors and topics provides more informative results than starting from the authors and linking the semantics of their communications.

Based on actor-network theory, and the emerging ideas about socio-semantic-network analysis, we focus on the connections between actors and topics mentioned in Twitter messages in two example cases. Different from social-network analysis, our focus is not on authors sending tweets to other authors, but on the co-occurrences of actors and topics, 'actants' in online debates.

*Whole matrix approach*

We have operationalized the whole matrix approach as follows. Each tweet is considered as a unit of analysis to which both actors and semantics (words) are attributed. The resulting matrix is asymmetrical, but one can generate an affiliations matrix of both words and actors in a single pass (by multiplication with the transposed of the matrix). The 2-mode matrix of words versus actors is contained in this matrix off-diagonal, whereas the co-word and co-author matrices are positioned along the main diagonal (Figures 1A and 1B).

The matrix in Figure 1A is similar to the word/document matrix as used in library and information science (Salton & McGill, 1983) and also widely used in social-network analysis (Borgatti & Everett, 1997) and recently also in semantic network analysis (e.g. Yang & González-Bailón, 2017). Figure 1B is a 1-mode matrix containing the semantic network of actors and topics, and their relations in a single representation. We call the latter, Figure 1B, the whole matrix approach.

|  | word 1 | word 2 | … | word m | actor 1 | actor 2 | … | actor k |
|---|---|---|---|---|---|---|---|---|
| tweet 1 |  |  |  |  |  |  |  |  |
| tweet 2 |  |  |  |  |  |  |  |  |
| tweet 3 |  |  |  |  |  |  |  |  |
| tweet 4 |  |  |  |  |  |  |  |  |
| tweet 5 |  |  |  |  |  |  |  |  |
| …. |  |  |  |  |  |  |  |  |
| tweet n |  |  |  |  |  |  |  |  |

**Figure 1A**: "Word/document" matrix of words and actors as attributes to tweets.

|  | word 1 | word 2 | … | word m | actor 1 | actor 2 | … | actor k |
|---|---|---|---|---|---|---|---|---|
| word 1 |  |  |  |  |  |  |  |  |
| word 2 |  | Semantic map |  |  |  | 2-mode network |  |  |
| … |  |  |  |  |  |  |  |  |
| word m |  |  |  |  |  |  |  |  |
| actor 1 |  |  |  |  |  |  |  |  |
| actor 2 |  | 2-mode network |  |  |  | Social network |  |  |
| … |  |  |  |  |  |  |  |  |
| actor k |  |  |  |  |  |  |  |  |

**Figure 1B**: Co-occurrence matrix of words and actors as the whole matrix approach

We argue that the results of the whole matrix provide more informative results than those based on the 2-mode matrix, in the case of socio-semantic-network analysis. We will discuss the similarities and differences of the 2-mode and the whole matrix approach in Results.

**Twitter data**

We chose to focus on Twitter because it provides the users with the option to tag their tweets as belonging to specific topics by using #hashtags, and addressing other users by @username. These two functions available at Twitter enable us to operationalize our approach. We discuss the implications for using other types data in the Discussion section below.

In general, Twitter enables users to send short, maximally 140-character messages to other Twitter users – and recent upgrading to allow for a maximum of 280-characters. The social media allows for addressing specific other users with adding the marker @ before the username of the targeted other user, retweeting messages authored by other Twitter users, for example by using the mark RT in the beginning of the message, and for tagging the messages using hashtags (with # mark) as well as spreading links to websites (using https;//t.co/url). These Twitter specific technological affordances (Foot & Schneider, 2006) allow for automated analysis of the Twitter messages—and the Twitter-specific functions. We discuss earlier findings related to the use of hashtags and usernames below.

Hashtags (e.g., Bruns and Stieglitz, 2012; Perez-Altable, 2015-2016,; Holmberg & Hellsten, 2016) and hashtags in combination with keywords (boyd, Golder & Lotan, 2010; Himelboim *et al*, 2017) have been used for selecting a data set for the analysis, and as identifying *ad hoc* publics on Twitter (Bruns & Burgess, 2010 ). Boyd *et al*. (2010) showed that 36% of tweets contained a @username and as few as 5% contained a #hashtag, whereas the more recent results by Gerlitz and Rieder (2013) presented 57,2% containing @usernames and 13% containing one or more hashtags. Saxton *et al*. (2015) manually coded the type of hashtags used by advocacy organizations, and found that tweets containing hashtags used by several type of organizations were more likely to be retweeted. Less research has focused on how different types of actor, such as non-governmental organizations, private individuals, and political parties use hashtags. Yet, Enlin and Simonsen (2017) show that politicians use significantly larger numbers of hashtags in their tweets than journalists. Bruns and Steiglich (2012) show that hashtags are used more often in original tweets, that are not retweets nor replies to other users. Hashtags are also more often used in relation to major media events, such as royal weddings or the rewarding of Oscar Prizes (Bruns & Stieglitz, 2012).

Earlier research has often focused on analyzing either co-occurring hashtags (e.g. Russell *et al*, 2010; Gerlitz & Rieder, 2013) or co-occurring usernames in tweets (Ausserhofer & Maireder, 2013; Pearce *et al*., 2014), but less on how these two co-occur in Twitter messages. On the use of usernames, Thelwall and Cugelman (2016) propose a resonating topic method for evaluating the success of campaigns by the United Nations Development Programme (UNDP), and found that usernames are used in relation to mentioning others in the tweets as well as replying to other users, in particular in connection to the RT@ format. We call both functions of using @usernames *addressing* other Twitter users.

We test the method on two data sets that differ in terms of 1) size of the data set, 2) language used in the tweets, and 3) the type of discussion, in order to validate the approach. Our *large-scale data set* consists of more than 100,000 tweets sent during the Rio+20 meeting in Rio de Janeiro, Brazil in the end of June 2012. This data set was collected using the open software crawler Webometric Analyst using the search term "#Rio+20" (Thelwall, 2009).[1] The data can be opened in Excel, and also includes a column for the language of the Twitter messages. We used this language column to select all English-language Twitter messages for our analysis. Out of the total 100,073 Twitter messages sent between 19 June and 2 July, 2012, 75,710 were in English. We further focus on the English-language tweets sent during the meeting, 20 to 22 June, 2012. This resulted in a data set of 72,077 tweets that were further analyzed. In this sample data set 5,211 unique usernames and 3,150 hashtags were mentioned in the Twitter messages.

---

[1] The first author is grateful to Prof. Mike Thelwall for collecting the data set in 2012 – only after this new method, it has become possible to analyze the Rio+20 tweets in a meaningful way.

Our *smaller data set* of Twitter messages was collected using the software tool Coosto from the period of 1 June 2015 to 1 June 2017 using the search term "vogelgriep AND pluimvee" ("birdflu AND poultry"). We downloaded 2,139 Twitter messages that include 234 unique @usernames and 230 unique #hashtags. The data set is in Dutch, but we discuss the results in English.

**Methods**

We developed two dedicated computer programs—tweet.exe and frqtwt.exe—that are available at https://leydesdorff.github.io/twitter. Frqtwt.exe reads a file (named "text.txt") containing the data and provides a word frequency distribution ("wordfrq.txt"). The analysis does not require the use of a stopwordlist for data cleaning since all the usernames and hashtags can be considered meaningful. By sorting the resulting file wordfrq.txt alphabetically, one can distinguish between words in the content of the messages, @usernames, and #hashtags. An advantage of frqtwt.exe is that it keeps the hashtags usernames and words in the messages as separate types of words. Thereby, the user can also easily remove the #hashtags and @usernames if the focus is on the words used in the Twitter messages, or as in our case, keep those in the analysis when the focus is specifically on hashtags and/or usernames. Alphabetical ordering of the words results in #hashtags positioned on the top of the word frequency list, followed by

@usernames. One can select the hashtags and the usernames to two separate files for setting respective thresholds if so wished.

Second, the routine tweet.exe reads the so generated file "words.txt" in combination with "text.txt" and generates the matrices shown in Figures 1A and B above. The resulting co-occurrence matrix of documents (tweets) versus words (hashtags and usernames) can be visualized using the open access softwares: Pajek (e.g. de Nooy *et al.,* 2012), and VOSViewer (van Eck & Waltman, 2007; 2011). Third, to visualize the results, and compare the results of the 2-mode and the whole matrix approaches, we used Pajek for constructing the network visualization files using Kamada-Kawai (1989) for the information layout, that we then exported to VOSViewer for the socio-semantic network visualizations.

**Results**

We discuss first the results using the small data set on bird flu and poultry in the Netherlands, and thereafter the results using the large-scale data set of Twitter messages sent during the Rio+20 environmental meeting in 2012. In both cases, we first discuss the similarities between the 2-mode and the whole matrix approach, and thereafter highlight the differences between the two analyses. In the end, we will show a further application of the method that results in a 3-mode network of Twitter authors (usernames sending the messages) as an additional layer to the co-addressed hashtags and usernames in the Rio+20 case.

*Bird flu tweets*

Bird flu, alias avian influenza has affected poultry farming, but also occasionally caused epidemics with human infections, most prominently in 2005-2006 when the H5N1 avian influenza virus spread from poultry to humans in Asia. The spread of the bird flu gained wide attention in the newspapers and online discussion forums (Hellsten & Nerlich, 2006). Recently, different strands of the avian influenza virus have infected poultry farms in Europe causing poultry farms to keep their poultry inside as well as regulations to temporarily stop and restrict the import of chicken from infected areas and transport of poultry for human consumption. We focus on Twitter discussions concerning bird flu in poultry in the Netherlands during the period 2015- 2017. There were two peaks in the number of tweets during this period, in December 2015 related to new cases of the disease in poultry farms in France, and in November-December 2016, related to cases in the Netherlands.

For the analysis of the co-occurring hashtags and usernames, we set the threshold to hashtags and usernames that were used five times or more often in the data set, for pragmatic reasons of keeping the number of nodes in the resulting visualizations roughly to 100 nodes. Using our programs, the user is free to set this threshold lower or higher depending on a specific research question, the size of the data, and the purpose of the study. For example, one may be interested in the diversity of hashtags and take samples of very frequently and very seldom-used hashtags and /or usernames, and compare across case studies.

**Figure 2A**: 2-mode matrix of 39 hashtags (green) and 63 usernames (red) used ≥5 times in 2,139 Twitter messages on "birdflu and poultry"; largest component contains 47 'actants'; VOSviewer used for the layout.

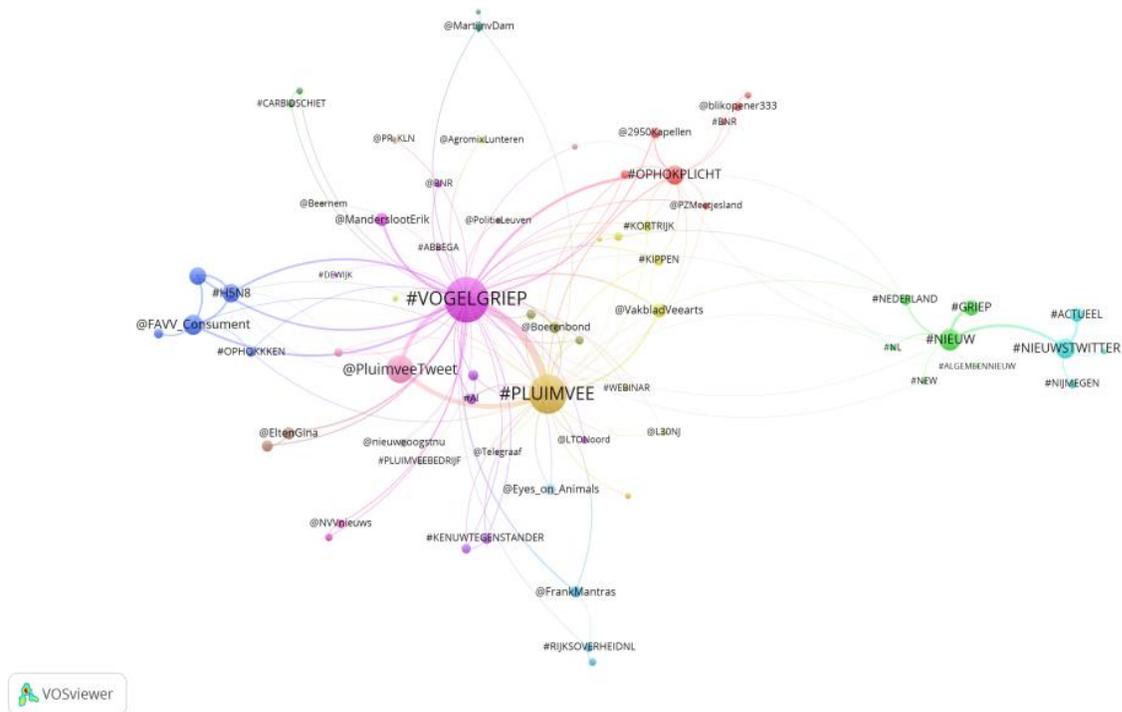

**Figure 2B:** Whole matrix of 39 hashtags and 63 usernames used ≥5 times in 2,139 Twitter messages on "birdflu and poultry"; largest component contains 67 'actants'; visualization: VOSviewer used for the layout and clustering.

Both Figures 2A and 2B show the main hashtags #vogelgriep and #pluimvee located central in the network together with the main organization that is targeted in the tweets @pluimveeTweet. The latter is an online newsfeed designated to poultry farmers. Updates of the situation are often retweeted in the data set, for example in the tweet about new regulations that will take place the next day:

```
RT @DNPPROVANT: Update vogelgriep: maatregel gaat in vanaf morgen
@FAVV_Consument https://t.co/PVMhT9p6fX
```

In most tweets, both the hashtags vogelgriep ("bird flu") and pluimvee ("poultry") are used together so that the tweet is tagged for both issues. For example, the news feed PluimveeTweet was first to send out the Twitter message on new cases of H5N1 epidemics in France, using both of the most common hashtags in the same tweet:

```
Hoogpathogene H5N1 #vogelgriep vastgesteld in Frankrijk
https://t.co/PdJzjwScqK #pluimvee
```

The map consists of a relatively large number of online news media (e.g. @PluimveeTweet, "poultryTweet" @GriepTweets, "fluTweets" and @LandbouwNieuws, "agricultureNews") and municipalities (#Kapellen, #Deerlijk, Heist-op-den-Berg, #Nijmegen) affected by the bird flu at poultry farms. Both Figures also show the same clusters around FAVV_Consument that is related to the Belgian Federal Agency for the Safety of Food, as well as the main regulations #ophokplicht and #ophokken ("indoor containment of the poultry").

However, the 2-mode visualization (Figure 2A) loses this regional clusters of hashtags, such as Nijmegen, a Dutch city located in the province of Gelderland as connected to #NieuwsTwitter, another online newsfeed (separate cluster on the left-hand side) and #griep ("flu") and #nieuws ("news") in Figure 2B. This is related to that 2-mode visualization cuts-off clusters consisting of only a single type of nodes – hashtags in our case, and hence lacks tweets such as:

```
#Nijmegen Landelijke maatregelen vogelgriep alleen nog voor
pluimvee, water- en loopvogels https://t.co/TaMs30tisW
#nieuwstwitter
```

This example above spreads information about national regulations for poultry, waterfowl and flightless birds in Nijmegen, tagging both the city of Nijmegen and one of the main newsfeeds,

NieuwsTwitter. The 2-mode analysis (Figure 2A) loses 20 actants when compared with the whole matrix (Figure 2B).

. The types of discussions (e.g. crisis, a summit, long term policy debate etc.) may result in different types of hashtag-username networks. Other types of actors can be prominent in other discussions. The map also shows non-governmental organizations active in environmental issues, such as Eyes_on_Animals, concerned with the effects of bird flu on food production. Such organizations, are positioned in the periphery of the map due to their lesser role in the Twitter discussion on bird flu. Our method provides an analytical tool to inspect how different types of actors are co-occurring with hashtags in addition to focusing on how specific authors use hashtags.

The results can be used in crisis management to identify national, regional and local online newsfeeds used by different organizations and citizens on Twitter for information spreading. In comparison, the whole matrix approach also shows clusters of one type of node, hashtags while 2-mode approach cuts those off the main component. The whole matrix approach performs more inclusively than the 2-mode approach.

*Rio+20 tweets*

To further validate the method, we use a large data set of tweets sent during the United Nations Conference on Sustainable Development, Rio+20 meeting, that took place in 2012. This meeting is also called the Earth Summit, and the RioPlus20 meeting as it took place twenty years after the Rio 1992 meeting on biodiversity conservation and climate change. The tweets sent during the Rio+20 meeting consist of a wide variety of participants discussing with each other during the meeting (e.g. locations of lunch meetings, general reporting during the speeches and about the meeting in general), media sending out live information during the meeting, and political bodies trying to influence the public opinion. This provides us a large data set of more than 72,000 tweets during a short-term event that we expect to consists of a high diversity of sub-topics discussed. Since the data was collected with the search term #RioPlus20, all the tweets contain per definition this hashtag; we removed this hashtag from the analysis (see Figure 3A and 3B).

**Figure 3A.** 2-mode matrix of the 47 hashtags (green) and 58 usernames (red) used ≥150 times in the 72,077 English-language Twitter messages sent during the Rio+20 meeting on 20-22 June, 2012: Largest component of 103 'actants' in the visualization, VOSviewer for the layout.

**Figure 3B**: Whole matrix of the 47 hashtags and 58 usernames used ≥150 times in the 72,077 English-language Twitter messages sent during the Rio+20 meeting on 20-22 June, 2012: Largest component of 104 'actants' in the visualization, VosViewer for the layout and clustering.

Whereas the largest connected component was of different size in the bird flu and poultry case, in the Twitter messages sent during the Rio+20 meeting, both approaches result in virtually the same number of 'actants' included in the visualizations. This is likely to be caused by the higher threshold used for the words (150 times or more often) for the analysis. We will discuss this further in relation to Figure 4 below.

In both the 2-mode and the whole matrix visualization (Figure 3A and 3B), one of the most prominent hashtags is #futurewewant; it is clearly present in the visualization. This hashtag connects several main actors during the meeting, such as @UN and @UNNewscenter. As an example, the hashtag has been used to retweet a message by WWF Australia and co-hashtagged with the general term @RioPlus20:

```
RT @WWF_Australia: .@UN_Rioplus20 We want a game changing set of
commitments tht will ensure a future w food, water & energy for all
@#futurewewant #RioPlus20
```

Both maps also show several sub-topics, for example around energy issues (#energy, #energyforall and @SGEnergyforall) and on women (#womenrio, @UNwomen). Global environmental NGOs, such as Oxfam, Greenpeace, and World Wildlife Foundation (WWF) are present in both visualizations. The NGO Greenpeace has also been co-addressed with a major newspaper, @guardian.

```
@Greenpeace moves to 'war footing' at #RioPlus20
http://t.co/nGjExgrN via @guardian
```

Both maps also show a strong activist cluster around the #endfossilfuelsubsidies linked to the actors @Avaaz and @dilmabr, the latter being the username of the former President of Brazil (on the right side in Figure 3A, and on the left in Figure 3B).

```
RT @Avaaz: You can find photos from our #EndFossilFuelSubsidies
activities on Facebook: http://t.co/2qJ0Lcre & Flickr
http://t.co/HXgNck4x #RioPlus20
```

However, the 2-mode matrix loses the connection between @Avaaz and @dilmabr in Figure 3A. Similar to the bird flu and poultry case above, this is caused by cutting-off connections between the same type of nodes, in the Rio+20 case between two @usernames.

*Adding authors to hashtag-username networks*

The analysis can be further refined by, for example, selecting tweet authors who have frequently posted on the issue, and then focusing on the co-occurring usernames and hashtags in the tweets by a specific active Twitter user, or organization. This further refining is particularly useful in the case of large and heterogeneous data sets, such as the Twitter messages during an international meeting. As an example, we therefore selected tweets that were sent out by two, different types of organizations, that sent out more than 150 tweets during the three-day meeting in Rio. Alternative way of doing this is to add the authors as an additional part to the right in the whole matrix (Figure 1b). The selected organizations are Greenpeace that sent out 173 tweets (combined from its different Twitter username accounts, such as Greenpeace_de, Greenpeace_UPA, GreenpeaceCA and GreenpeaceNZ), and the Asian Development Bank (ADB) that sent out 160 tweets in our data set (combined from the different local Twitter user accounts of the bank, such as ADB_Manila, ADBandNGOs, ADBClimate and ADBEnvironment). The 173 tweets sent by Greenpeace during the three-day meeting used 15 unique hashtags and 15 unique usernames twice or more times whereas the 160 tweets sent out by ADB consist of 30 unique hashtags and 20 usernames used twice or more often. For both authors we included these hashtags and usernames addressed in the tweets that were used at least two times (Figure 4).

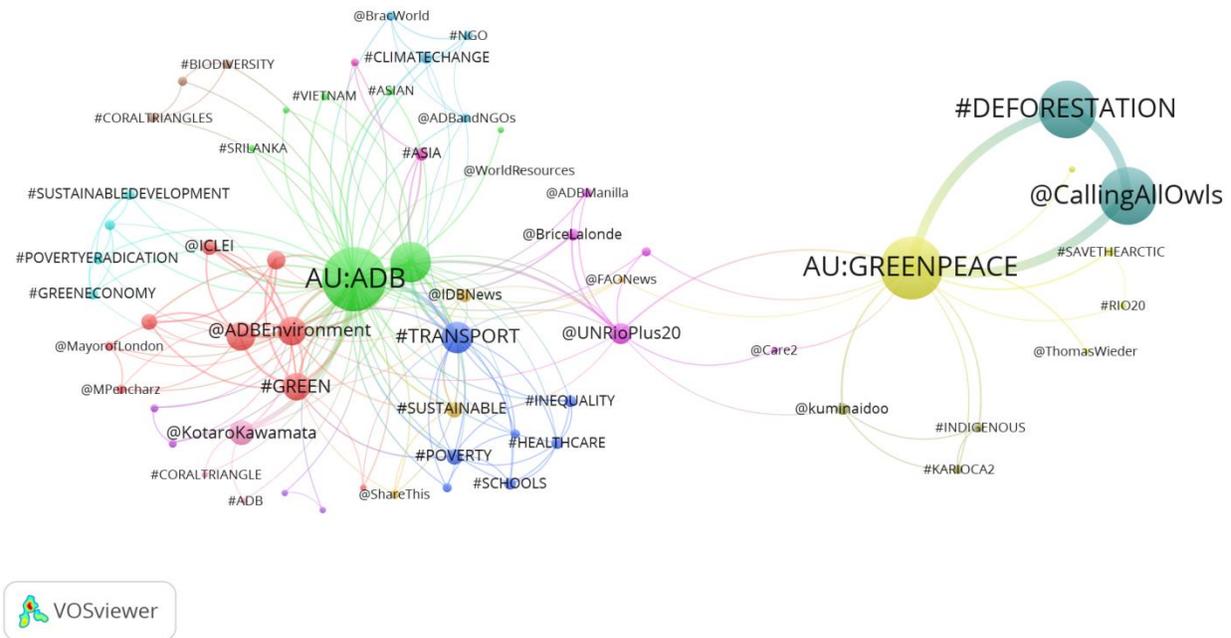

**Figure 4:** 3-mode network of the two main organizations (Greenpeace and ADB) as 'authors', and the hashtags and usernames addressed in their 173 and 160 tweets during the Rio+20 meeting. Main component of 61 'actants' and 2 'authors', VOSViewer for the layout and clustering.

Figure 4 shows that the two very active organizations (in terms of the number of tweets sent), Greenpeace and the Asian Development Bank (ADB) mainly participated in their own sub-debates during the meeting. The main shared hashtag is #futurewewant that is also central in Figures 3A and 3B above. Both organizations also refer to shared usernames, such as @UNRioPlus20 and @FAONews.

Greenpeace was mainly co-addressing the topics of #RioPlus20 and #deforestration, linked with the username @CallingAllOwls that refers to a campaign of painting owls to save forest in order to pressure towards zero deforestation by 2020. A typical tweet sent by Greenpeace is shown below:

```
Greenpeace is @CallingAllOwls - pls RT and @ it to leaders
#RioPlus20 + Zero #deforestation. One of 1000  voices:
http://t.co/K9WiD5R0
```

Interestingly, the main hashtag addressed by Greenpeace, #deforestration, remained isolated in the context of all the tweets sent during the Rio+20 meeting (Figure 3B on the left side) which indicates that the campaign was not highly retweeted by the other Twitter users during the meeting

ADB, in turn, was involved in several topical discussions, such as #poverty, # inequality, #healthcare and (lower left-hand side), and #greeneconomy #sustainabledevelopment (right-hand side):

```
Poor #transport exacerbates #poverty and #inequality, inhibiting
access to #schools, #healthcare, markets & job opportunities.
#rioplus20
```

The results show a more detailed view on the activities of the selected organizations in participating in the debate on Twitter. As an advantage of further splitting the data according to the authors of the tweets is that one can compare across different actor types; for example, due to the smaller size of the data we are able to include hashtags and usernames that were use twice or more often, in the network visualization.

In summary, the method can be extended into 3-mode or even higher-order network analyses because it takes into account the whole matrix, as presented in Figure 1b. This is an improvement as compared to the 2-mode approach. In particular, the whole matrix approach performs better for socio-semantic-network analysis where the two types of nodes are co-addressed in texts. This is because it includes only clusters consisting of a similar kind of nodes.

Noteworthy, in 2012, the mark @ was used not only in combination with a username to address another user but also to designate a location, simply replacing the word 'at':

```
RT @makower: Ted Turner @ UN Foundation dinner: "Clean coal:
Bullshit." #rioplus20
```

Our method is able to differentiate between these two uses of the @ symbol: one can manually change or remove the @ sign referring to location from the data set. In the 2015-2017 data set the mark @ was solely used in combination with a username, as a conventional way to address other Twitter users. Perhaps, this indicates changes in the use of social media tools over time. More research is needed to analyze in detail how the use of other social media tools, beyond Twitter, has evolved over time. Such developments pose new challenges for social scientists interested in longitudinal studies of social media content. We discuss more implications of the whole matrix approach in the Discussion and Conclusion section.

**Discussion and Conclusion**

We have proposed a new methodological approach for analyzing Twitter messages by focusing on the co-occurrences of Twitter specific #hashtags and @usernames instead of the words used in the content of the Twitter messages. Our approach has the advantage to enable the mapping of which users were addressed in connection to which topics. This approach helps to solve the problem of semantic networks that have been criticized of producing "bags-of-words" that remain vague in terms of meaningful interpretations. We have shown the advantage of the whole matrix approach in providing more complete results than the 2-mode approach, in particular, in also including clusters that consist of either hashtags or usernames. 2-mode matrix tends to cut-off such clusters. In addition, the whole matrix approach allows for extending the analysis from two types of nodes into *n*-mode networks ($n > 2$). As an example, we extended the analysis to a 3-mode network of authors, actors and hashtags, and mapped the results in a single visualization (Figure 4). This adds opportunities for researchers to focus on multiple types of nodes depending on their research questions.

To theory-building, mapping hashtags and usernames instead of the words used in the message content provides a more informative overview of the online discussions – co-occurrences of specific actors related to hashtags provides information on which actors were addressed in relation to which topics, hence advancing actor-network theory by, indeed, analyzing hashtags and actors as 'actants' based on their connections (Latour, 1995). In the context of actor-network theory (Latour 2005, Callon *et al.*, 1986), the results are first steps toward automating the analysis of socio-semantic networks using text documents, in a way that does not rely on social networks between authors. In our approach we show, instead, the connections between actors and topics in online discussions. As our approach does not require focusing on the most active Twitter users, we are able to account for relations of addressing actors and the topics as co-occurring 'actants'. Further theory-building for the implications of our empirical research is needed.

To the emerging field of socio-semantic networks, previously applied to both offline (Saint-Charles & Mongeau, 2017; Basov *et al.*, 2017) and online communications (Roth, 2013; Roth & Cointet, 2000), our approach proposes a new empirical way of studying small as well as large scale data sets in a way that provides meaningful results of the co-addressed 'actants' in the communications. To our knowledge, this is the first automated effort to investigate how actors and topics are co-addressed in mediated communications.

Furthermore, our approach provides an improvement to the 2-mode approach that has been applied to social-network analysis as the main methodological approach since the 1990s (Borgatti & Everett, 1997). Whereas the 2-mode approach has proven fruitful for the analysis of bi-graphs, for example, of authors and words, the whole matrix approach seems to perform more inclusive for analysis by combining actors and topics. There is need for further theoretical and methodological research into comparing the two approaches with different types of data sets.

In practical terms, one of the additional advantages of this approach is that it works without data cleaning such as removing plural forms of the words, stemming of words, and using a stopword list to remove less meaningful words (e.g. the, a, an, he, she, it etc.) from the analysis. All hashtags and usernames are meaningful without any need for cleaning. The routines are also not limited with the size of the data set, in our case they were applicable to a smaller data sets of a few thousand tweets and to a data set of more than hundred thousand tweets. This allows for a more reliable bottom-up approach to social media discussions.

In conclusion, our approach can be applied to a wide range of theoretical traditions in the communication sciences, such as research into issue arenas as well as stakeholder analysis by focusing on the co-mentioning of actors in news media, social media, and organizational media in general. Although we applied the method to the Twitter messages under study, the approach can also be applied to, for example, scientific publications where subject headings or keywords—meta-data—can be considered as #hashtags and actors cited in the texts as @usernames. An additional future research can combine social-network analysis of the relations between the authors of the tweets and those targeted in the tweets. As a further step, the approach can also be used for analyzing other types of texts by visualizing for example the organization names addressed in newspaper articles, similar to @username in Twitter messages. Alternatively, the approach can also be used for scientific texts using subject categories or keywords as #hashtags and mentioned author names as @usernames.

More research is needed to further validate and improve the method, and to find optimal ways to apply it including meta-data, that lack the use of # and @ markers in the texts. This empirical research can feedback into theory-building in the information and communication sciences that marks a shift from actor-based approaches to content-based approaches in network analysis.